\documentclass[12pt]{article}

\usepackage{epsfig,amsfonts,amssymb,newlfont,rotate,float,array}

\DeclareMathAlphabet{\mathitbf}{T1}{cmr}{bx}{it}

\renewcommand{\d}{\mathrm d}
\newcommand{\e}{\mathrm e}
\newcommand{\IDelta}{\mathit{\Delta}}
\newcommand{\bbox}[1]{\mbox{\boldmath $#1$}} 

\begin{document}
\title{The four dimensional site-diluted Ising model: a finite-size
scaling study}
\author{H.~G.~Ballesteros\footnote{\tt hector@lattice.fis.ucm.es}~,~
        L.~A.~Fern\'andez\footnote{\tt laf@lattice.fis.ucm.es}~,\\
        V.~Mart\'{\i}n-Mayor\footnote{\tt victor@lattice.fis.ucm.es}~,~
        A.~Mu\~noz Sudupe\footnote{\tt sudupe@lattice.fis.ucm.es}~,\\
\normalsize \it Departamento de F\'{\i}sica Te\'orica I, 
        Facultad de CC. F\'{\i}sicas,\\  
\normalsize \it Universidad Complutense de Madrid, 28040 Madrid, Spain.\\
\\
        G.~Parisi\footnote{\tt giorgio.parisi@roma1.infn.it} ~and~
        J.~J.~Ruiz-Lorenzo\footnote{\tt ruiz@chimera.roma1.infn.it}~.\\
\normalsize \it Dipartimento di Fisica and Istituto Nazionale di 
        Fisica Nucleare,\\ 
\normalsize \it Universit\`a di Roma ``La Sapienza'', P.~A.~Moro  2, 
        00185 Roma, Italy.
}

\date{July 16, 1997}

\maketitle

\thispagestyle{empty}

\vglue -5mm\vskip -5mm

\begin{abstract}

Using finite-size scaling techniques, we study the critical properties
of the site-diluted Ising model in four dimensions . We carry out a
high statistics Monte Carlo simulation for several values of the
dilution. The results support the perturbative scenario:
there is only the Ising fixed point with large logarithmic 
scaling corrections. 
We obtain, using the
Perturbative Renormalization Group, functional forms for the scaling
of several observables that are in agreement with the numerical data.
\end{abstract}

Monte Carlo.
Disordered Systems.
Critical exponents.
Finite size scaling.
$\epsilon$-expansion.


\newpage

\section{\protect\label{S_INT}Introduction}

A possible way to obtain new Universality Classes (UC) is to add disorder to
known pure systems. The Harris criterion~\cite{HARRIS} says that, if
the specific heat diverges with a power law in the pure system, then,
the disorder will change the critical behavior of the model, i.e.
a new UC will appear. Conversely, if the specific heat does not diverge
in the pure system, then, the critical exponents of the disordered
system will remain unchanged. In the limiting case, what amounts to
a discontinuous or a logarithmically divergent specific heat, the
criterion does not apply and we need to study analytically or/and
numerically the system.

The quest and the characterization of new UC are very important in
dimensions two and three (with a direct relevance on Condensed Matter
Physics) and in four dimensions (with implications on High Energy
Physics).
In the last case it is crucial to characterize all the possible
UC, in order to be able to define a Field Theory
on a non-perturbative basis. As the Gaussian model gives a trivial 
one (i.e. at long distances the theory will be free) 
we are interested in finding a non-Gaussian UC.

In this paper we will study the four dimensional site-diluted Ising model
that was previously studied~\cite{PARU} by two of the authors, who
calculated numerically the critical exponents, analyzing the divergences with
the temperature. Their results pointed to non
Gaussian critical exponents, for large values of the dilution, but it
was noted the possibility of a crossover between the behavior found
and the Gaussian one.

In order to obtain accurate measures of the critical properties we
have repeated the simulations in a greater number of spin configurations.
The use of Finite Size Scaling (FSS) techniques allows us to work in
large lattices at the critical point.

Using the Perturbative Renormalization Group (PRG) equations we
calculate the dependence of the observables at the critical point with
the scale of the system including logarithmic corrections. These are
different from the pure system.

Our determination of the critical exponents and other critical
properties matches very well with the predictions of the PRG: a Gaussian
critical behavior with logarithmic corrections.
We check this behavior along the critical line in a wide range of
concentrations, from $p=0.8$ to $p=0.3$ (the percolation threshold is
near 0.2).  We remark that a scenario based on hyperscaling seems
completely unlikely from our numerical simulations.

The structure of the paper is as follows. In the next section we
define the model and the observables. In section 3 we present our
analytical calculations, obtaining the FSS formulas and calculating
the values of two different Binder cumulants in the thermodynamic
limit. In section 4 we describe the numerical methods and the
different techniques that we will use to analyze the observables. In
section 5 we show our numerical results confronting them with the
analytical predictions. Finally we report the conclusions.

\section{\protect\label{MODEL} The model}

The model we study numerically in this article, is defined in terms of
Z$_2$ spin-variables placed in the nodes of a hypercubic
four-dimensional lattice. The action is

\begin{equation}
S=-\beta\sum_{<i,j>} \epsilon_i \epsilon_j \sigma_i \sigma_j\ ,
\label{HAMILTONIANO}
\end{equation}
where the sum is extended over nearest-neighbors, and the
{$\epsilon_i$} are quenched, uncorrelated random variables, taking the
value $1$, with probability $p$, and $0$ with probability $1-p$. An
actual \{$\epsilon_i$\} configuration, will be called a {\it sample}
from now on. For every observable it is understood that first one
performs the Ising model calculation and then the $\epsilon$-average.

In the following, we shall denote an Ising average with brackets,
while the sample average will be overlined. The observables will be
denoted with calligraphic letters, i.e.  $\cal O$ and with italics the
double average $O=\overline{\langle\cal O\rangle}$. We define the
total nearest-neighbor energy and the normalized magnetization as

\begin{equation}
{\cal E} =\sum_{\langle
i,j\rangle}\epsilon_i\sigma_i\epsilon_j\sigma_j\ , \qquad {\cal
M}=\frac{1}{V}\sum_i \epsilon_i\sigma_i\ ,
\end{equation}
$V$ being the volume (defined as $L^4$, where $L$ is the lattice
size). We also define the susceptibility as

\begin{equation}
\chi=V\overline{\left\langle {\cal M}^2 \right\rangle}\ .
\end{equation}
Another very useful quantity is the Binder parameter:

\begin{equation}
g_4=\frac{3}{2}-\frac{1}{2}\frac{\overline{\langle {\cal M}^4\rangle}}
           {\overline{\langle {\cal M}^2 \rangle^2}}\ .
\end{equation}
Other kind of Binder parameter, meaningless for the pure system, can be
defined as
\begin{equation}
g_2=\frac{\overline{ \langle {\cal M}^2 \rangle^2 - 
\overline{\langle {\cal M}^2 \rangle}^2
}}
{\overline{ \langle {\cal M}^2 \rangle}^2 } \ .
\end{equation}

A very convenient definition of the correlation length in a finite lattice, 
reads~\cite{XIL}

\begin{equation}
\xi=\left(\frac{\chi/F-1}{4\sin^2(\pi/L)}\right)^\frac{1}{2},
\label{XI}
\end{equation}
where $F$ is defined in terms of the Fourier transform of the
magnetization
\begin{equation}
{\cal F}(\mathitbf{k})=\frac{1}{V}\sum_{\mathitbf{r}}\e^{\mathrm i
\mathitbf{k}\cdot\mathitbf{r}} \epsilon_{\mathitbf r}\sigma_{\mathitbf
r}\ ,
\end{equation}
as
\begin{equation}
F=\frac{V}{4}\overline{\left\langle |{\cal
F}(2\pi/L,0,0,0)|^2+\mathrm{permutations}\right\rangle}\ .
\end{equation}

This definition is very well behaved for the finite-size scaling (FSS) method 
we employ~\cite{OURFSS}, and it is also fair natural for
considerations about  triviality~\cite{TRIVIAL}. 
Finally, we measure the specific heat
\begin{equation}
C= V^{-1} \overline{\langle\cal E^2 \rangle  -\langle\cal E \rangle ^2} \ .
\end{equation}

\section{\protect\label{S_THB}Analytical predictions}

The field-theoretical study  of the model~(\ref{HAMILTONIANO})
is performed
by means of a $\phi^4$ theory with a random mass term, whose action is
\begin{equation}
S[\phi]=\int \d^d x~  \left( \frac{1}{2} 
\left(\partial_\mu \phi\right)^2 +
\frac{1}{2} m^2(x)  \phi^2 + \frac{1}{4!} v \phi^4   \right)
\ .
\end{equation}

Here it is assumed that the mass term is a quenched,
spatially-uncorrelated, stochastic variable. We will later argue that
the only relevant parameters of the disorder distribution are its
mean, $r$, and variance, $\IDelta^2$, so we will assume for simplicity the
Gaussian distribution

\begin{equation}
\d P[m^2(x)]\propto \d[m^2] \exp\left(- \frac{(m^2(x)-r)^2}{2
\IDelta^2}\right) \ . 
\end{equation}

When the disorder is quenched we need to compute, in a first step, the
free energy of the system for a given choice of the disorder (in this
case of the mass term), and then average this free energy with the
probability distribution of the disorder.  To manage this kind of
problems it is very useful to use the so-called replica
trick~\cite{GIORGIO1}.

Let us introduce $n$ replicas of the initial system, 
$\phi_i$, with $i=1,\ldots, n$. The average of the replicated
partition function over the Gaussian disorder will be denoted by
overlines.

\begin{equation}
F=\overline{\log {\cal Z}}=\lim_{n \to 0} \frac{1}{n}
\left(\overline{{\cal Z}^n}-1
\right)
\ .
\end{equation}

Now we can define an effective action by means of

\begin{equation}
\overline{{\cal Z}^n} = 
{\cal Z}_{\mathrm{eff}}=\int \d[\phi_i]~ \exp(-S_{\mathrm{eff}}[\phi_i]) \ ,
\end{equation}
with
\begin{equation}
\label{action}
\!S_{\mathrm{eff}}[\phi_i]=\!\int\! \d^d x \left[ \frac{1}{2} \sum_{i=1}^n 
\left(\partial_\mu \phi_i\right)^2 +
\frac{r}{2} \sum_{i=1}^n \phi_i^2 +\frac{u}{4!} \left[\sum_{i=1}^n
 \phi_i^2\right]^2 + \frac{v}{4!} \sum_{i=1}^n  \phi_i^4 \right]\!,\!
\end{equation}
where $u=- 3 \IDelta^2$. This gives us a starting
point for the analytical calculation. The $n \to 0$  
limit should be taken at the end.

For $v=0$ the action is $O(n$)-invariant. When $u=0$ the action
describes $n$ decoupled Ising models. We remark that $u$ is negative
and proportional to the dilution. It is possible to show that for a
non-Gaussian distribution, terms associated with higher connected
momenta of the distribution appear in the effective action.  The
$s$-momentum couples to $\phi^{2s}$ thus, if $s>2$, is irrelevant in
four dimensions and can be neglected.  In our numerical simulation a
site is occupied with probability $p$, so $\IDelta^2 = p(1-p)$.

The action (\ref{action}) was studied in ref.~\cite{AHARONY} by 
using PRG techniques. Considering a differential
dilatation, the following equations are obtained:
\begin{eqnarray}
\nonumber
\frac{\d r}{\d \log b} &= & 2 r+4 K_d (2u+3 v) (1-r) \ ,\\ 
\label{rg_eq}
\frac{\d v}{\d \log b} &= &\epsilon v - 12 K_d v (4 u + 3v) \ , \\ 
\nonumber
\frac{\d u}{\d \log b} &= &\epsilon u - 8 K_d u (4 u + 3 v) \ ,
\end{eqnarray}
where $\epsilon \equiv 4-d$, $d$ being the dimension, $K_d$ is a
constant that depends on the dimension, and $b$ is the Renormalization
Group (RG) scaling factor.  In the previous formulas third order terms
in $u,v$ have been neglected, and the $n\to 0$ limit has been
considered.  In the following we shall set
$\epsilon=0$, as we study a four dimensional problem.

The initial conditions of the system, typically verify $|u_0| \ll
v_0$. Therefore, the RG
evolution, driven by eqs.~(\ref{rg_eq}), will present two interesting
regimes

\begin{enumerate}
\item A transient regime in which $-u \sim v^{2/3}$. We find $v(b)
\sim 1/\log b$ or equivalently $v(L) \sim 1/\log L$.
\item An asymptotic regime reached by following the RG evolution until
the equation $4u+3v=O(u^2)$ finally holds. We obtain, including the
next term in the perturbative expansion, $u^2(b),v^2(b) \propto 1/\log
b$, or $ 1/\log L$.
\end{enumerate}

Thus, we expect a crossover between the pure situation where
the relevant coupling, $v$, goes to zero as $v(L) \sim 1/\log
L$ and the disordered one where $u$ and $v$ are similar in magnitude
and $v(L)\sim 1/\sqrt{\log L}$, i.e. the relevant coupling ($u$ or $v$)
goes to zero more slowly than in the initial regime.

Defining $t \equiv r-4 K_4 u$, 
eqs. (\ref{rg_eq}) in the asymptotic regime reduce to

\begin{eqnarray} \nonumber
\frac{\d t}{\d \log b}&=&2 t +8 K_4 u t \ , \\
\frac{\d u}{\d \log b}&=&-\frac{1696}{3} K_4^2 u^3 \ .
\end{eqnarray}

For large $b$, the solutions are 

{\setlength{\extrarowheight}{-8pt}
\begin{equation}
\begin{array}{rcl}
t(b) &=& t_0 b^2 
\exp \left[-2 \sqrt{\displaystyle\frac{3 \log b}{53}} \right] \ , \\
\\
u(b) &=& - \sqrt{\displaystyle \frac{3}{3392 K_4^2 \log b} }\ ,
\label{UDEB}
\end{array}
\end{equation}
}where $t_0$ is an integration constant and in the large $b$ regime $u(b)$ 
does not depend on the initial condition $u_0$.

Using these formulas it is possible to obtain the expressions for the
correlation length, susceptibility and specific heat as functions of
the reduced temperature.

We find just one slight difference with
ref.~\cite{AHARONY}.
The equation for the wave function renormalization, $\zeta(b)$, is
\begin{equation}
\frac{\d \zeta}{\d \log b}=-\gamma_\phi(u,v)=-8 K_4^2 (2 u^2 +6 u v +3 v^2)\ .
\end{equation}
In the pure model $\zeta$ is constant, but in the
asymptotic region one obtains \hbox{$\zeta(b) \propto (\log
b)^{1/212}$} which affects to the susceptibility. This correction is
negligible from a practical point of view, but it could be important
in related models (for instance in spin glasses and
percolation~\cite{JJRL}). So the formula for the susceptibility reads:
\begin{equation}
\chi \simeq t_0^{-1} 
\exp\left[\sqrt{\frac{6}{53}} |\log t_0|^{1/2} \right]
|\log t_0|^{1/106} \ ,
\end{equation}
where we have used that $\chi=\zeta^2~\xi^2$~\cite{LE_BELLAC}. In
ref.~\cite{AHARONY} the term $\zeta^2$ is absent.

\subsection{\protect\label{S_BINDER}Calculation of 
Binder Cumulants}

In this section we will calculate the values of $g_2$ and $g_4$, with
the techniques introduced in ref.~\cite{ZINN} for the study of
finite geometries using Field Theoretical methods.

The main idea is to expand the field $\phi(x)$ in Fourier modes. 
In a finite geometry the biggest contribution comes from the
zero mode. It can be shown that it has to be treated non
perturbatively while this is not necessary for the rest of the
modes~\cite{ZINN}.

In our case the effective action for the zero mode, that we will
denote as $\psi_i$ is, in a $L^d$ volume  and just at the MF 
critical point (i.e. $r=0$),
\begin{equation}
\label{action_mf}
S_{\mathrm{eff}}[\psi_i]= L^d \left[ \frac{u}{4!} \left(\sum_{i=1}^n
 \psi_i^2\right)^2 + \frac{v}{4!} \sum_{i=1}^n  \psi_i^4 \right] .
\end{equation}
and the partition function is
\begin{equation}
{\cal Z}_\mathrm{eff}(n)=\int \left(\prod_{i=1}^n \d \psi_i 
\right) \exp(-S_{\mathrm{eff}}[\psi_i]) \ .
\end{equation}

In the asymptotic regime, the relation $4 u +3 v\simeq 0$ is satisfied
with good precision, and so:

\begin{equation}
\label{action_mf2}
{\cal Z}_\mathrm{eff}(n)=
\frac{1}{\sqrt{3 \pi}} \int \left(\prod_{i=1}^n \d \psi_i 
\right) \d \lambda ~ \exp\left[-\frac{1}{3} \lambda^2 +\lambda \sum_i \psi_i^2
-\sum_i \psi_i^4\right] \ , 
\end{equation}
where we have introduced a Gaussian $\lambda$-integration 
in order to decouple the
term $(\sum_{i=1}^n \psi_i^2)^2$. It is possible to see that
dimensionless ratios, like $g_4$ and $g_2$, do not depend on the specific
value of $v$, thereby we have also fixed $v=4!/L^d$ in the previous
formula and in the rest of the section. 

We remark that the ratio between $u$ and $v$ in $d\le 4$ is universal,
because there, we have a (limiting) fixed ratio $u/v$.  For instance,
in four dimensions $u$ and $v$ go to zero with a limiting ratio, $u/v
\to -4/3$, whereas in $d<4$, $u$ and $v$ go to the non trivial fixed
points $u^*$ and $v^*$, respectively, and $u/v \to u^*/v^*$. In $d >
4$ it is impossible to fix this ratio: it depends on the parameters in
the Hamiltonian.  It is possible to show that using the
$\epsilon$-expansion one can obtain for a Binder cumulant the MF value
(calculated in $d > 4$) plus corrections that are proportional to real
and positive powers of $\epsilon$~\cite{ZINN}. So we can fix the ratio
between $u$ and $v$ to the four dimensions value and then do the
computation directly in $d > 4$ (i.e. avoiding the loop effects).

We can perform the integrals
on the $\psi$ variables
\begin{equation}
{\cal Z}_\mathrm{eff}(n)= 
\frac{1}{\sqrt{3 \pi}} \int \d \lambda ~ \e^{-\lambda^2/3} I_0(\lambda)^n \ ,
\end{equation}
where
\begin{equation}
I_m(\lambda) \equiv \int \d \psi \exp\left[ \lambda \psi^2-\psi^4
\right] \psi^m \ .
\end{equation}

For the calculation of the cumulants we need to evaluate with the 
action (\ref{action_mf2}) the following averages~\cite{GIORGIO1}:
\begin{eqnarray}\nonumber
\overline{\langle {\cal M}^2 \rangle} & \rightarrow & \langle \psi_a^2 \rangle
\ , \\    
\overline{\langle {\cal M}^4 \rangle} & \rightarrow & \langle \psi_a^4 \rangle
\ , \\    \nonumber
\overline{ \langle {\cal M}^2  \rangle^2} & 
        \rightarrow & \langle \psi_a^2 \psi_b^2 \rangle
\, \,\mathrm{with}  \,\, a \neq b \ .    
\end{eqnarray}

For instance
\begin{equation}
\langle \psi_a^{2m} \rangle =
(\sqrt{3\pi}\cal Z_\mathrm{eff})^{-1}{\int \d \lambda ~
e^{-\lambda^2/3} I_{2m}(\lambda) I_0(\lambda)^{n-1} }\ .
\end{equation}

The moments in the $n \to 0$  limit  are

\begin{equation}
\begin{array}{rcl}
\langle \psi_a^{2m} \rangle &=& 
\displaystyle\frac{1}{\sqrt{3 \pi}} \int \d \lambda
\ \frac{I_{2m}(\lambda)}{I_0(\lambda)}\ \e^{-\lambda^2/3}\ , \\
\langle \psi_a^2 \psi_b^2 \rangle &=& 
\displaystyle\frac{1}{\sqrt{3 \pi}} \int \d
\lambda  \left[\frac{I_2(\lambda)}{I_0(\lambda)}\right]^2
\e^{-\lambda^2/3} \ ,
\label{integrals}
\end{array}
\end{equation}
where $a \neq b$.

Evaluating numerically the previous integrals we obtain
\begin{eqnarray}
g_4^\mathrm{disordered}&=&0.32455\ldots \ ,\\
g_2^\mathrm{disordered}&=&0.31024\ldots \ . 
\label{G4G2}
\end{eqnarray}
We recall
that the MF values for the moments in the pure case are~\cite{ZINN}
\begin{equation}
\langle  {\cal M}  ^{2m}  \rangle= \frac{I_{2m}(0)}{I_0(0)} \ ,
\label{mf}
\end{equation}
and for the cumulants
\begin{eqnarray}
g_4^\mathrm{pure}&=&0.40578\ldots\ ,\\
g_2^\mathrm{pure}&=&0\ .
\end{eqnarray}
Now the interpretation of the formulas (\ref{integrals}) is clear. We
have a  $\lambda$-model with action 
$S= \lambda \psi^2 -\psi^4$, with $\lambda$
distributed with a Gaussian weight
$\exp(-\lambda^2/3)$. Averaging with this probability
distribution the moments  $I_{2m}(\lambda)/I_0(\lambda)$ 
of these $\lambda$-models
we obtain the right Binder cumulants for the diluted one.  Obviously
when the mass term is zero we recover the MF result~(\ref{mf}) for the
pure model.

\subsection{\protect\label{S_FSS}FSS in the diluted model}

The scaling of the singular part of the free
energy in the presence of a magnetic field, $h_0$, is
\begin{equation}
\label{main_rg}
f_{\mathrm{sing}}\left(r_0,u_0,v_0,h_0,\frac{1}{L}\right) 
=b^{-4} f_{\mathrm{sing}}\left(r(b),u(b),v(b),h(b),\frac{b}{L}\right) \ ,
\end{equation}
where we have introduced a new coupling, the system size $L$, which
scales trivially with a RG transformation. 
As usually the magnetic field verifies~\cite{LE_BELLAC}:
\begin{equation}
\frac{\d \log h(b)}{\d \log b} =\frac{d}{2}+1
-\frac{\gamma_\phi(u,v)}{2}\ .
\label{RGH}
\end{equation}

In the asymptotic regime, the solution of (\ref{RGH}) is
\begin{equation}
h(b)= h_0 b^3 (\log b)^{\frac{1}{212}}\ .
\label{HDEB}
\end{equation}

Performing a RG transformation with $b=L$, we keep just one degree of
freedom (see ref.~\cite{BLOTE} for more details).  The free
energy of this system in the asymptotic regime is 
\begin{equation}
\begin{array}{lcr}
f(r',u',h',L=1) \equiv
\displaystyle\log \int \left(\prod_{i=1}^n \d \phi_i \right)\times\\
\displaystyle\exp\left\{-\left[\frac{r'}{2} \sum_{i=1}^n \phi_i^2 
- h'\sum_{i=1}^n
\phi_i+ \frac{u'}{4!}\left(\sum_{i=1}^n
 \phi_i^2\right)^2  -
 \frac{u'}{3!} \sum_{i=1}^n \phi_i^4 \right]\right\}\ .
\end{array}
\end{equation}
We re-scale the $\phi_i$ variables by means of
$\phi_i^\prime=u^{1/4} \phi_i$. The free energy can be written as
\begin{equation}
\label{second_rg}
f(r',u',h',L=1)= 
{\hat f}\left(\frac{r'}{u'^{1/2}},1,\frac{h'}{u'^{1/4}}\right)\ ,
\end{equation}
obtaining finally
\begin{equation}
f_{\mathrm{sing}}\left(r_0,u_0,v_0,h_0,\frac{1}{L}\right)=L^{-4} 
{\hat
f}\left(\frac{r(L)}{u(L)^{1/2}},1,\frac{h(L)}{u(L)^{1/4}}\right)\ .
\label{FSINGFHAT}
\end{equation}

We remark that the $u$ variable is a {\em dangerous} (marginally)
irrelevant variable ~\cite{SOKAL2,FISHER}, and we need to do with care
all the analytical steps (it is not correct to substitute $u$ for its
asymptotically value, $u=0$, because the free energy depends on
inverse powers of $u$).

As $\hat f$ is an analytical function, the $n\to 0$ limit can be
taken in the RHS of (\ref{FSINGFHAT}) substituting $r,h,u$ by
their limiting values. For clarity, in the following we will omit the
$L$ dependence in the functions $r,u$ and $h$.

To compute the thermodynamical quantities in the critical region one
just need to take the appropriate derivatives of $f_\mathrm{sing}$.
It will prove convenient to keep in mind  
eqs. (\ref{UDEB}) and (\ref{main_rg}), and the
following Taylor expansion 
(which depends on the relations $r=t+4 K_4 u$
and the fact that $t(L)=0$ whenever $t_0=0$)
\begin{equation}
\begin{array}{rcl}
\left. \partial^2_i{\hat f}(r/u^{1/2},1,0)\right|_{t_0=0}&=&
\partial^2_i{\hat f}(4 K_4 u^{1/2},1,0) \\
\\
&= & \displaystyle\partial^2_i{\hat f}(0,1,0) +O(u^{1/2}) \ ,
\end{array}
\label{taylor}
\end{equation}
where $\partial_i$ is the partial derivative with respect to the
$i$-th argument.

As we are interested in the behavior with the
lattice size just at the infinite volume critical temperature, the
susceptibility can be written as
\begin{eqnarray}\nonumber
\chi &\propto& \left. \frac{\partial^2 f_\mathrm{sing}}
{\partial h_0^2}\right|_{h_0=t_0=0}
=L^{-4} 
\left(\frac{\partial h}{\partial h_0}\right)^2 \left. 
\frac{\partial^2}{\partial h^2}{\hat f}(r/u^{\frac{1}{2}},1,h/u^{\frac{1}{4}})
\right|_{h_0=t_0=0} \\
&\simeq& L^2 (\log L)^{\frac{1}{4}+\frac{1}{106}}\ .
\label{sus}
\end{eqnarray}

The specific heat can be computed analogously
\begin{eqnarray}\nonumber
C &\propto& \left. \frac{\partial^2 f_\mathrm{sing}}
{\partial t_0^2}\right|_{h_0=t_0=0} 
=L^{-4} 
\left(\frac{\partial r}{\partial t_0}\right)^2
\left.\frac{\partial^2}{\partial r^2} {\hat
f}(r/u^{\frac{1}{2}},1,h/u^{\frac{1}{4}})\right|_{h_0=t_0=0} \\
&\simeq&  (\log L)^{\frac{1}{2}}\e^{-\sqrt{\frac{48}{53}|\log L|}}
\ .
\end{eqnarray}

At zero magnetic field, the correlation length scales as
\begin{equation}
\left. \xi(r_0,u_0,1/L) \right|_{t_0=0}=L \left. \xi(r,u,1) 
\right|_{t_0=0} \ ,
\label{XI1}
\end{equation}
where $\xi(r,u,1)$ must be evaluated with the free energy
(\ref{second_rg}). Consequently, the mass squared term is
\begin{equation}
\left( \left.\xi(r,u,1)\right|_{t_0=0}\right)^{-2}
= \left. \frac{r}{u^{1/2}}
\right|_{t_0=0} \propto u^{1/2} \ ,
\label{XI2}
\end{equation}
and so
\begin{equation}
\xi(r_0,u_0,1/L) \propto \frac{L}{u^{1/4}} \simeq L (\log L)^{\frac{1}{8}} \ .
\label{XIFIN}
\end{equation}

Finally we can also compute the shift of the apparent critical
temperature. It can be defined  as the temperature where the
susceptibility (or specific heat)  measured in a finite volume 
shows a maximum. Using the
formula (\ref{sus}) for the susceptibility
without imposing the constraint
$t_0=0$  we obtain
\begin{equation}
\chi \propto L^2 (\log L)^{\frac{1}{4}+\frac{1}{106}}
\partial^2_3{\hat f}(r/u^{1/2},1,0)\ .
\end{equation}

The maximum of $\chi$ as a function of $L$ and $t$ is not just at
$t_0=0$, but it is fixed by the condition  
\begin{equation}
r/u^\frac{1}{2}=(t + 4 K_4 u)/u^\frac{1}{2}=x_\mathrm{max} \ ,
\end{equation}
i.e. the function
$\partial^2_3{\hat f}(x,1,0)$ has a maximum at $x=x_\mathrm{max}$ .

As  \hbox{$t \propto T_\mathrm{c}(\infty)-T_\mathrm{c}(L)$}, it follows that
\begin{equation}
T_\mathrm{c}(\infty)-T_\mathrm{c}(L) \propto L^{-2} (\log L)^{-\frac{1}{4}} 
\e^{\sqrt{\frac{12}{53} \log L}} \ .
\label{shift}
\end{equation} 

To finish this section we will report, for completeness, the finite
size formulas for the same observables in the pure case~\cite{KENNA}:
\begin{eqnarray}\nonumber
\xi&\propto&L(\log L)^{\frac{1}{4}}     \ ,       \\ \nonumber
\chi&\propto&L^2(\log L)^{\frac{1}{2}}  \ ,        \\
C &\propto& (\log L)^{\frac{1}{3}} \ ,           \\
T_\mathrm{c}(\infty)-T_\mathrm{c}(L) 
&\propto & L^{-2} (\log L)^{-\frac{1}{6}}\nonumber \ .
\label{FSSPURO}
\end{eqnarray}

The latter expressions can also be obtained with the method described
above.

\section{\protect\label{S_NM}Numerical Methods}

The choice of
the Monte Carlo (MC) update algorithm should be carefully
considered. The Wolff single cluster method~\cite{WOLFF} is the best
choice for the pure model. However in the diluted case, small
isolated clusters of spins are very likely to appear. Those clusters
are scarcely visited by the Wolff method. Therefore, we have
complemented the updating method with a Metropolis sweep per
measure. We have checked that this algorithm thermalizes appropriately
the configurations for $p\geq 0.5$, by comparing the numerical results
from cold and hot starts. However, for $p\leq 0.4$, equilibration by
this method becomes really hard to achieve. This is due to
the presence of intermediate size clusters almost isolated from the
percolating one.  We have then turned to the Swendsen-Wang
(SW)~\cite{SW} algorithm which guarantees that all-sized spin-clusters
are considered. In this way, we find complete agreement between hot
and cold starts. We have also compared the results of a SW and a
single-cluster simulation at $p=0.5$ on our largest lattice, finding
compatible results. However, to get a statistically-independent new
configuration takes significantly longer (in CPU time) with the SW
algorithm.

For every observable, one first averages in the sample, then
averages between different samples. Therefore, a crucial point
is how long each sample simulation should be. Assuming full statistical
independence between different measures (quite possible with a cluster
method), and also between measures taken in different samples, the variance
of such a mean is 

\begin{equation}
\sigma^2_T=\frac{1}{N_S}\left(\sigma^2_S+\frac{\sigma^2_I}{N_I}\right)\
,
\label{ERROR}
\end{equation}
where $N_S$ is the number of samples generated and $N_I$ is the number of
measures in each Ising-model simulation. The variance between samples
of the thermal average of our observable is $\sigma^2_S$ and, finally,
 $\sigma^2_I$ is the average of the variances in each
sample. On the other hand, the
computational effort is roughly proportional to $N_S N_I$, as the
computer usually spends a fixed fraction of the time measuring. It is
then clear that the optimum value of $N_I$ cannot be much bigger that
$\sigma^2_I/\sigma^2_S$. One could even be tempted to measure just
once by sample. However, we shall come back to this point when
discussing re-weighting methods.

\subsection{Derivatives and Re-weighting Methods}

\begin{figure}[t]
\begin{center}
\leavevmode
\centering\epsfig{file=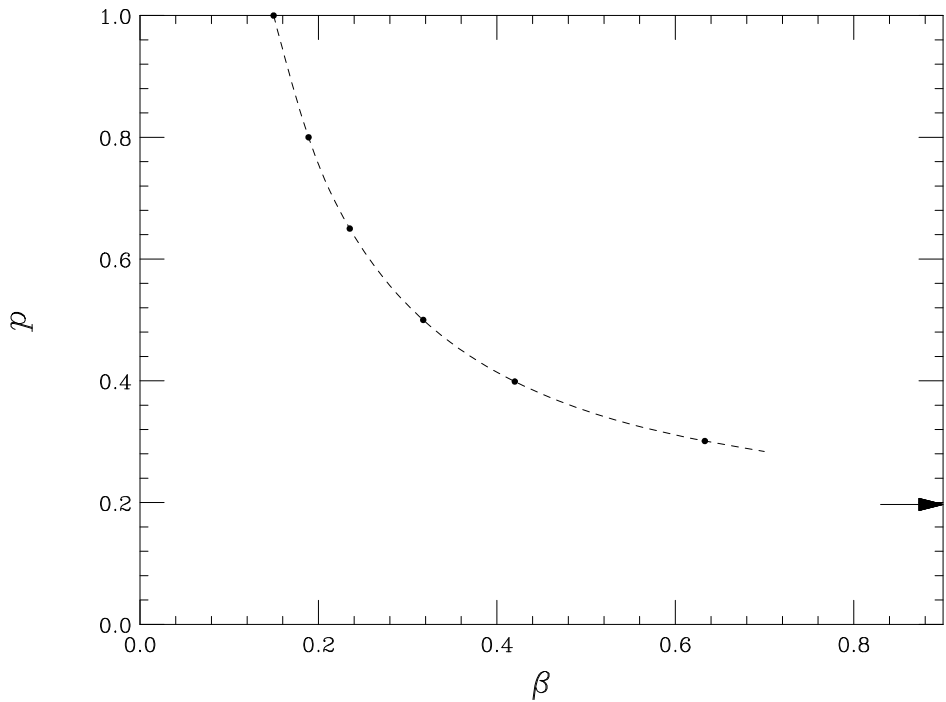,width=0.68\linewidth,angle=90}
\end{center}
\protect\caption{Phase diagram of the four dimensional site-diluted 
Ising model. The points correspond to the simulated values, and the
arrow indicates the percolation limit.
\protect\label{PHASES}}
\end{figure}

The critical curve slope (see fig.~\ref{PHASES}) changes quite
abruptly.  It is nearly vertical for large $p$ and almost horizontal
close to the percolation threshold. It is therefore wise to choose a
re-weighting method such that we may extrapolate to different $\beta$
values close to the pure model, but to different $p$ values at very
strong dilution. Let us then comment both extrapolation methods
separately.

\subsubsection{$\bbox{\beta}$-extrapolation}

The energy measures allow the calculation of $\beta$-derivatives 
of observables, and
the use of the standard re-weighting methods, before the
sample-average is performed. However, an important point should be
made now, so let us recall how are they calculated:

\begin{equation}
\partial_\beta \overline{\langle \cal O\rangle}=
 \overline{\partial_\beta\langle \cal O\rangle}=
\overline{\left\langle_{\vphantom{|}} {\cal{OE}} - \langle\cal O
 \rangle \langle \cal E \rangle\right\rangle}\ ,
\label{DERIVADA}
\end{equation}
\begin{equation}
\langle \cal O \rangle (\beta+\Delta\beta)=
\langle \cal O \e^{\Delta\beta\cal E}\rangle /
\langle \e^{\Delta\beta\cal E}\rangle\ .
\label{FS}
\end{equation}

It is clear that both expressions are biased. For instance, the
expectation value of eq. (\ref{DERIVADA}), when the averages are
calculated with $N_I$ measures, is really
\begin{equation}
\overline{\left(1-\frac{2\tau}{N_I}\right)\partial_\beta \langle\cal O
\rangle}\ ,
\label{BIASDER}
\end{equation}
$\tau$ being the integrated autocorrelation time~\cite{SOKAL}, which
depends on the sample.
The bias for eq. (\ref{FS}) is also of order $2\tau/N_I$, but terms of
higher order in $1/N_I$ are to be expected. 

These biases are immaterial
for usual MC investigations, as the statistical error decreases with
the square root of the number of measurements. However, in our case
the bias is of order $1/{N_I}$, while the statistical error is of
order $1/\sqrt{N_S}$, which are similar in our
simulations! 

The cure for this is to introduce unbiased estimators. A possible
method would be to measure in completely independent samples,
ensuring $2\tau=1$ for every sample. 
In this case the unbiased estimator is constructed
multiplying by $1/(1-1/N_I)$  the
$\beta$-derivatives and by other more complex functions the extrapolated
observables. However, this would be too expensive from the computational
point of view. The solution we find is to work with $\tau\gtrsim 1$,
repeat the calculations with different values of $N_I$ and extrapolate
$N_I\to\infty$.

Specifically, in each sample, we calculate the derivative with the
full MC history, obtaining a number $y_1$, the bias being proportional
to $2\tau/N_I$.  We then consider two contiguous halves of the MC
history, repeat the calculation for each one and average the final
result. This value, $y_2$, has a bias that goes as $4\tau/N_I$. The
next term, $y_3$, is obtained with four quarters with a bias
proportional to $8\tau/N_I$.

The linear extrapolation is
\begin{equation}
y_L = 2 y_1 - y_2\ ,
\end{equation}
and the quadratic one 
\begin{equation}
y_Q = \frac{8}{3} y_1 - 2 y_2 + \frac{1}{3} y_3\ . 
\end{equation}

\begin{figure}[t]
\begin{center}
\leavevmode
\centering\epsfig{file=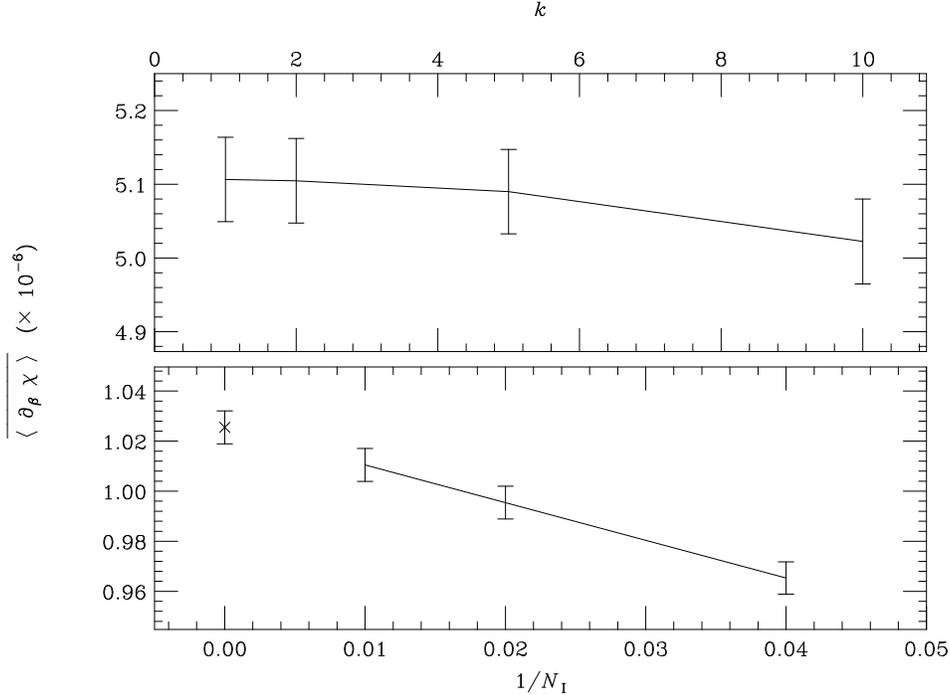,width=0.68\linewidth,angle=90}
\end{center}
\protect\caption{In the upper part we show the sample-averaged $y_1$
value for $\partial_\beta\chi$, taking one out of each $k$ measures,
in a $L=48$ lattice at $p=0.5$, with $\tau\approx 8$. As both $\tau$
and $N_I$ get divided by $k$, we obtain an stable value until
$k=10$. In the lower part, we plot the sample-averaged $y_1$, $y_2$
and $y_3$ values, as a function of the inverse number of MC measures,
in a $L=32$ lattice, at $p=0.5$, with $\tau\approx 0.8$. The linear
behavior is apparent.  \protect\label{TAU}}
\end{figure}

We then average $y_L$ and $y_Q$ for all the samples, checking that the
difference is negligible compared with the statistical error. In fact,
we have found that the slope in the $(y,1/N_I)$ plane is a very clean
and easy measure of $\tau$, which we have used to achieve statistical
independence between different measures. See fig.~\ref{TAU} for an
example.

We proceed analogously with the extrapolations 
of observables or their derivatives.

Another approach to eliminate the biases is to split the measures in
statistically independent sets and  multiply the average
of some operator in the first set by the average of another operator in
the second set. In some cases (for instance derivatives
extrapolated at different couplings) it could be necessary to work with
more than two independent sets.  As in practice one have contiguous MC
measures, the statistical independence of the measures becomes involved
when splitting in several sets. 

The comparison between the quadratic
extrapolation and the linear one makes it possible to monitor this
{\em short MC history} effects. In some cases due to the necessity of
a large extrapolation in the coupling, we have found  differences 
between $y_L$ and $y_Q$ around one half of the statistical error;
then we have repeated the simulation at a nearer point.

\subsubsection{$\mathitbf{p}$-extrapolation}

In addition to the standard $\beta$-extrapolation~\cite{REWEIGHT},
it is also possible to extrapolate the mean values obtained
at a given dilution probability, $p$, to a close one $p'$
(see~\cite{PERC}). Let us simply recall that the
probability of finding a occupation number $q$, when filling
sites with probability $p$, is binomial. Therefore the 
re-weighting is calculated from a set of $N_S$ mean values of an 
observable $\cal O$ and the actual density
of the configuration $\{(\langle\cal O \rangle_i(\beta),q_i)\}$:

\begin{equation}
 \overline{\langle\cal O\rangle} (p',\beta)=
\frac{1}{N_S}{\displaystyle\sum_i^{N_S} \left(\frac{p'}{p}\right)^{q_iV}
        \left(\frac{1-p'}{1-p}\right)^{(1-q_i)V} \langle\cal O
\rangle_i(\beta)}\ .
\label{REWEIG} 
\end{equation}

The visible region is of course constrained to the variance of the binomial
distribution $p(1-p)/V$. Fortunately, it has been enough for us.

Using equation (\ref{REWEIG}) $p$-derivatives of observables can also
be computed, but statistical errors are about eight times bigger than
for $\beta$-derivatives. Therefore, our choice is to study 
 $p$-extrapolated $\beta$-derivatives, where all the above comments for the
bias in the derivative are in order, but the re-weighting is unbiased.

\subsection{Numerical FSS techniques}

As already stated in the introduction, 
we consider the possibility of finding a non-Gaussian fixed point in
four dimensions, where hyperscaling relations were fulfilled.  In such
a case, the usual FSS ansatz is expected to hold. However, the
PRG analysis rather suggests the presence of logarithmic
corrections to the Gaussian behavior. Therefore, we should also
consider the modifications in the FSS ansatz induced by the
logarithmic corrections. In this section, we shall first remind 
how critical exponents are measured (see refs.~\cite{OURFSS,PERC} for
similar calculations),
then we shall show how to deal with logarithmic corrections.

When hyperscaling holds, a very accurate way of
measuring critical exponents involves a form of the FSS ansatz where
everything is directly measurable on a lattice:
\begin{equation}
O(L,\beta,p)=L^{x_O/\nu}\left(F_O(\xi(L,\beta,p)/L)+O(L^{-\omega})\right)\
,
\label{FSS}
\end{equation}
where a critical behavior $t^{-x_O}$ is expected for the operator $O$,
$\omega$ is the universal scaling-corrections exponent, and $F_O$ is a
(smooth) scaling function. Notice that terms of order
$\xi^{-\omega}_{L=\infty}$ are dropped from eq.  (\ref{FSS}), so we
are deep within the scaling region. From a Renormalization Group point
of view, $\omega$ corresponds to the leading irrelevant operator.

In order to calculate the critical exponents, we study  
the quotient of $O(sL)$ and $O(L)$, defined as
\begin{equation}
Q_O=O(sL,\beta,p)/O(L,\beta,p)\ .
\end{equation}

Measuring at a value of the couplings
where the quotient for the correlation-length is $s$  
the scaling function may be eliminated and we obtain:

\begin{equation}
\left.Q_O\right|_{Q_\xi=s}=s^{x_O/\nu}+O(L^{-\omega})\ .
\label{QUO}
\end{equation}

If there are logarithmic corrections to hyperscaling, the critical behavior 
of the operator $O$ is modified to
\begin{equation}
O(L,\beta_c,p_c)\propto L^{x_O/\nu} (\log L)^{\delta_O}\ .
\label{with_log}
\end{equation}
We remark that the scaling variable is again $\xi(L,\beta,p)/L$.
This is clear from
formulas (\ref{XI1}), (\ref{XI2}) and (\ref{FSINGFHAT}): the scaling
variable is $r/u^{1/2}$ that is equal to $(\xi/L)^2$, and so one
readily obtains that the scaling variable is $\xi/L$ without
logarithmic corrections.

For the susceptibility exponent we find from eq. (\ref{sus}) after
some algebra
\begin{equation}
\frac{x_\chi}{\nu}=\frac{ \log \left.Q_\chi\right|_{Q_\xi=s}}{\log s}
+ O\left(\frac{1}{\log L}\right) \ .
\end{equation}
Thus, the corrections to the $\eta$ exponent are proportional to 
$1/\log L$ as in the pure case.

To estimate the logarithmic correction to the critical behavior
of the $\beta$-derivative of the correlation length we start from
eqs. (\ref{XI1}) and (\ref{XI2}) and get
\begin{equation}
\xi=\xi(r_0,u_0,1/L)=L \frac{u^{1/4}}{r^{1/2}}\ .
\end{equation}
Taking the $t_0$ derivative and using eqs. (\ref{UDEB}) we obtain
\begin{equation}
\partial_\beta \xi \propto L^3  \left(\log L \right)^{1/4}
\left(\frac{\xi}{L} \right)^3 
\exp \left[-2 \sqrt{\frac{3 \log L}{53}} \right]\ .
\label{expo}
\end{equation}
It is easy to check that   
\begin{equation}
\frac{x_{\partial_\beta \xi}}{\nu}= 
\frac{ \log \left.Q_{\partial_\beta \xi}
\right|_{Q_\xi=s}}{\log s}
+ O\left(\frac{1}{\sqrt{\log L}}\right)\ ,
\label{NULOGS}
\end{equation}
where the correction $O(1/\sqrt{\log L})$ arises from 
the exponential term in eq.~(\ref{expo}). 

For the lattice sizes simulated, we expect some dependence on the
dilution in the coefficient of the $1/\sqrt{\log L}$ term. In the
initial regime the corrections are proportional to $1/\log L$ (the pure
model).  Until the system forget the initial conditions (in particular
the dependence on the dilution) the coefficient of the $1/\sqrt{\log
L}$ term could change.

\section{\protect\label{S_NR}Numerical Results}

The lattice sizes that we have studied have been $L=8,12,16,24$ and $32$.
We have generated $N_S$=10,000 samples, for each lattice size, at
dilution values $p=0.8,0.65,0.5,0.4,0.3$.  In each sample, we measure
$N_I=100$ times after equilibration. The number of clusters traced (or
SW updates) between measures have been chosen to yield $2\tau\approx
1$ (see eq. (\ref{BIASDER})). The pure ($p=1$) model has also been
studied for $L=8,12,16,24,32,48$ and $64$, as a
contrast of the disorder-induced effects.
 
We shall present our numerical results in two steps. First we shall
consider the conventional FSS analysis (i.e. assuming hyperscaling),
finding that the percolation scenario is {\it extremely} unlikely.

After that, we shall look for hyperscaling violations in the data. We
shall find that they can be measured, and are indeed of the same order 
as predicted by PRG.

\subsection{Assuming Hyperscaling}

To measure the critical exponents, we use the so-called quotients
method, which allows for a great statistical
accuracy~\cite{OURFSS,PERC}. The starting point is eq. (\ref{QUO}). We
have first approximately located the point where

\begin{equation}
\frac{\xi(2L,\beta,p)}{2L}=\frac{\xi(L,\beta,p)}{L}\ ,
\label{BETANAIVE}
\end{equation}
then we have used re-weighting techniques to fine-tune the
condition~(\ref{BETANAIVE}). For $p=1.0$, $0.8$, $0.65$ and $0.5$, we
have used, $\beta$-extrapolation. Therefore, the critical $p$ is
fixed, $\beta$ being the tunable parameter. For lower dilution values,
we have rather used $p$-extrapolation, so we have first approximately
located the $\beta$ value, for which condition~(\ref{BETANAIVE}) holds
at $p\simeq 0.3,0.4$. Next we fix $\beta$, fine-tuning $p$
afterwards, so the critical values differ from $p=0.3,0.4$ by an amount
of less than $1\%$. However, in tables and figures, we shall refer to
them as $p=0.4$ and $0.3$ for brevity.

\begin{figure}[t]
\begin{center}
\leavevmode
\centering\epsfig{file=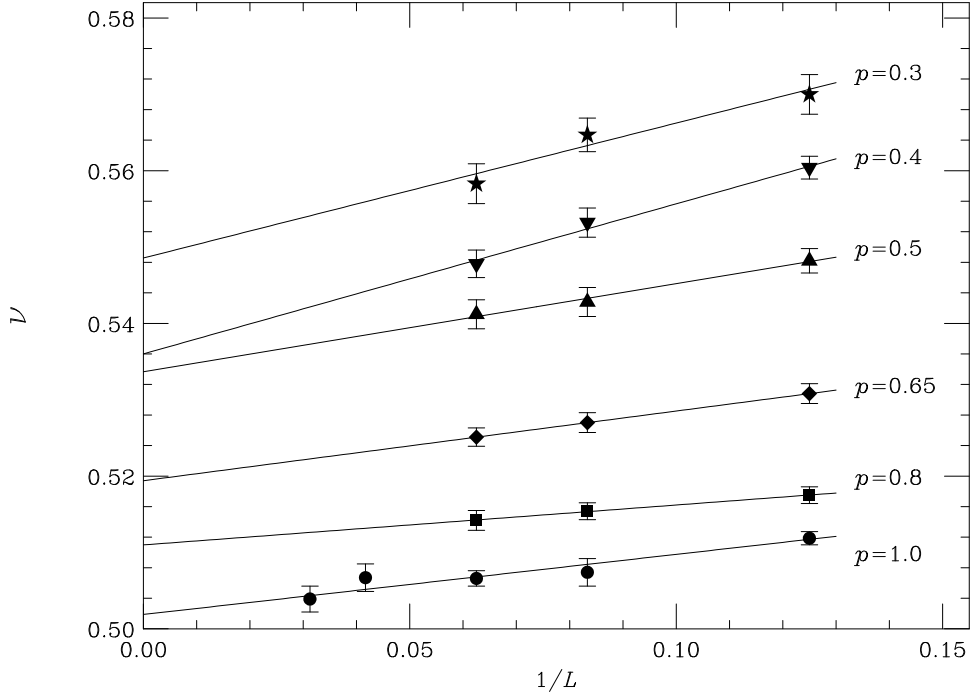,width=0.68\linewidth,angle=90}
\end{center}
\protect\caption{Exponent $\nu$ for different lattice pairs of type
$(L,2L)$. The lines correspond to linear fits.
\protect\label{NU1/L}}
\end{figure}

Eq.(\ref{QUO}) applied to the operators $\partial_\beta \xi$ and
$\chi$, yields respectively the exponents $1+1/\nu$ and $2-\eta$.  The
numerical results are shown in tables~\ref{NUNAIVE} and
\ref{ETANAIVE}. For the  $\nu$ exponent we find, instead of a stable
value, a monotonically decreasing one. For $\eta$, such an evolution
with growing $L$ is found, but it is clearly weaker. Therefore, an
infinite volume extrapolation is called for.  If hyperscaling holds,
we expect finite-volume scaling-corrections proportional to
$L^{-\omega}$.  As $\omega=1.13(10)$ in the percolation~\cite{PERC},
in the last row of both tables~\ref{NUNAIVE} and \ref{ETANAIVE} we include
an infinite-volume extrapolation with $\omega=1$. This fit is shown in
figure~\ref{NU1/L}.

\begin{table}[t]
\footnotesize
\begin{center}
\begin{tabular}{|r|l|l|l|l|l|l|}\hline

$L$&   \multicolumn{1}{c|}{$p=1.0$}      
    & \multicolumn{1}{c|}{$p=0.8$}      
    & \multicolumn{1}{c|}{$p=0.65$}      
    & \multicolumn{1}{c|}{$p=0.5$}      
    & \multicolumn{1}{c|}{$p\simeq 0.4$}      
    & \multicolumn{1}{c|}{$p\simeq 0.3$}\\ \hline\hline

8&0.5119(9)&0.5175(11)&0.5308(13)&0.5482(16)&0.5604(15)&0.5700(26)\\\hline
12&0.5074(18)&0.5154(11)&0.5270(13)&0.5428(19)&0.5532(19)&0.5647(22)\\\hline
16&0.5066(10)&0.5142(13)&0.5251(12)&0.5412(19)&0.5478(18)&0.5583(26)\\\hline
24&0.5067(18)&&&&& \\\hline
32&0.5039(17)&&&&& \\\hline
$\infty$&0.5019(14)&0.5110(25)&0.5194(26)&0.534(4)&0.536(4)&0.549(5)\\\hline
\end{tabular}
\caption{The $\nu$ exponent for  $(L,2L)$ pairs at the different
dilutions.}
\label{NUNAIVE}
\end{center}
\end{table}

\begin{table}[t]
\footnotesize
\begin{center}
\begin{tabular}{|r|l|l|l|l|l|l|}\hline
\multicolumn{1}{|c|} {$L$}

&   \multicolumn{1}{c|}{$p=1.0$}      
    & \multicolumn{1}{c|}{$p=0.8$}      
    & \multicolumn{1}{c|}{$p=0.65$}      
    & \multicolumn{1}{c|}{$p=0.5$}      
    & \multicolumn{1}{c|}{$p\simeq 0.4$}      
    & \multicolumn{1}{c|}{$p\simeq 0.3$}\\ \hline\hline

8& -0.02644(14)&-0.0161(12)&-0.0124(11)&-0.0055(19)&0.002(8) &-0.003(4)\\
\hline
12&-0.02132(26)&-0.0138(12)&-0.0089(13)&-0.0051(17)&0.000(9) &-0.0043(21)\\
\hline
16&-0.01656(12)&-0.0130(12)&-0.0053(10)&-0.0073(17)&0.000(11)&-0.006(7)\\
\hline
24&-0.01259(29)&&&&& \\\hline
32&-0.01445(18)&&&&& \\\hline
$\infty$&-0.0085(18)&-0.0097(25)&0.0013(22)&-0.008(4)&-0.002(20)&-0.007(9)
\\\hline
\end{tabular}
\caption{The $\eta$ exponent for $(L,2L)$ pairs at different dilutions.}
\label{ETANAIVE}
\end{center}
\end{table} 

It is clear that the percolation scenario, $\nu=0.686(2)$ and
$\eta=-0.094(3)$ \cite{PERC}, can be ruled out.  Moreover, the
possibility of a different fixed point neither Gaussian nor with the
percolation critical exponents requires a fairly exotic FSS behavior.
We do not find this possibility likely.

At this point, one could claim that this model presents {\it weak
universality}, as recently proposed in the two dimensional version of
the model~\cite{KIM}. That is, the exponent $\eta$ is constant over the
critical line, while $\nu$ is continuously varying. However, a much
less spectacular, but more likely interpretation will be given in the next
subsection. 

\subsection{The quest for logarithms}

\begin{figure}[t]
\begin{center}
\leavevmode
\centering\epsfig{file=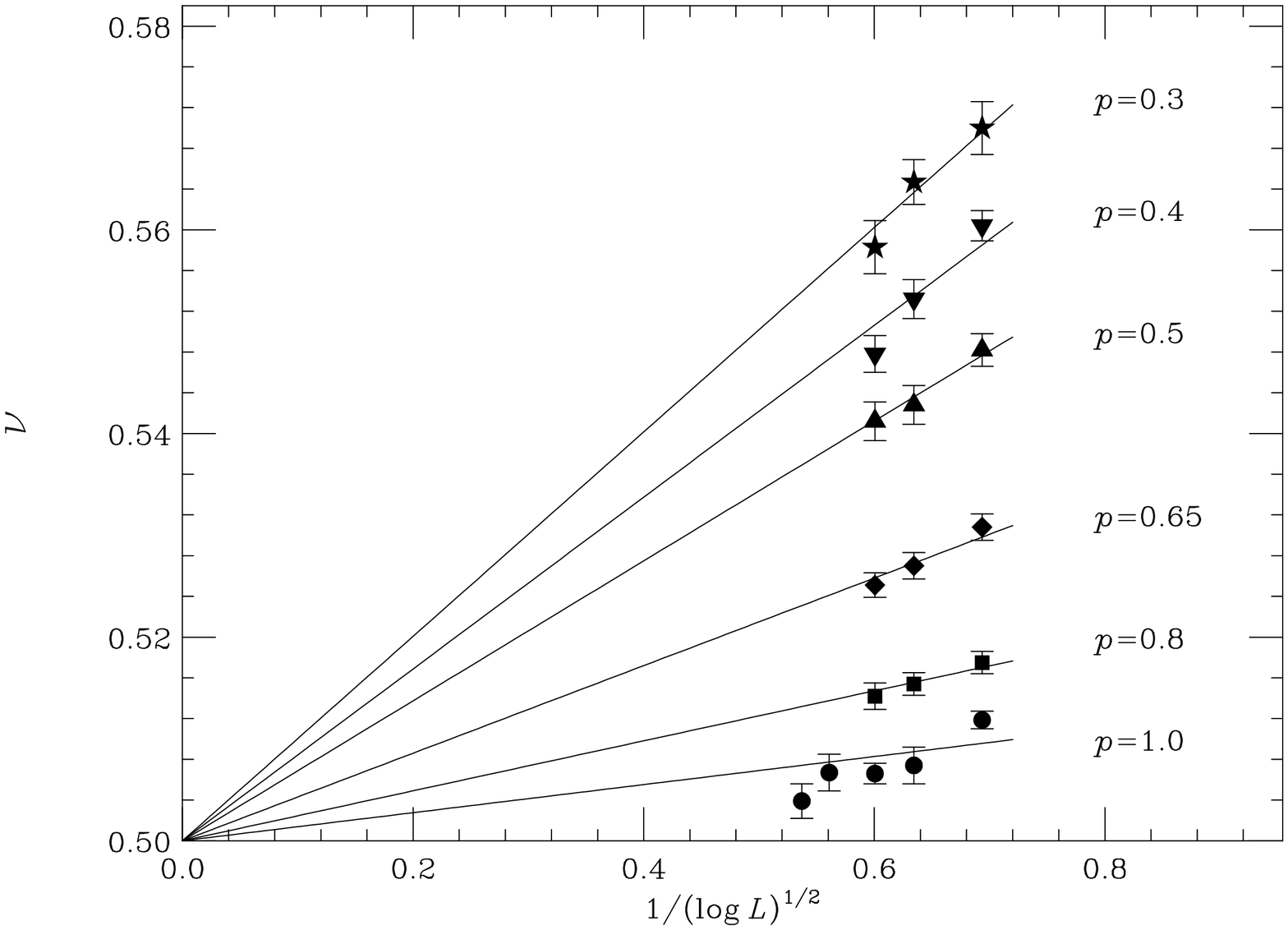,width=0.68\linewidth,angle=90}
\end{center}
\protect\caption{The $\nu$ exponent obtained from $(L,2L)$ pairs. The
solid lines are linear fits constrained to be $\nu=0.5$ in the $L\to\infty$
limit.
\protect\label{NU1/logL}}
\end{figure}

\begin{figure}[t]
\begin{center}
\leavevmode
\epsfig{file=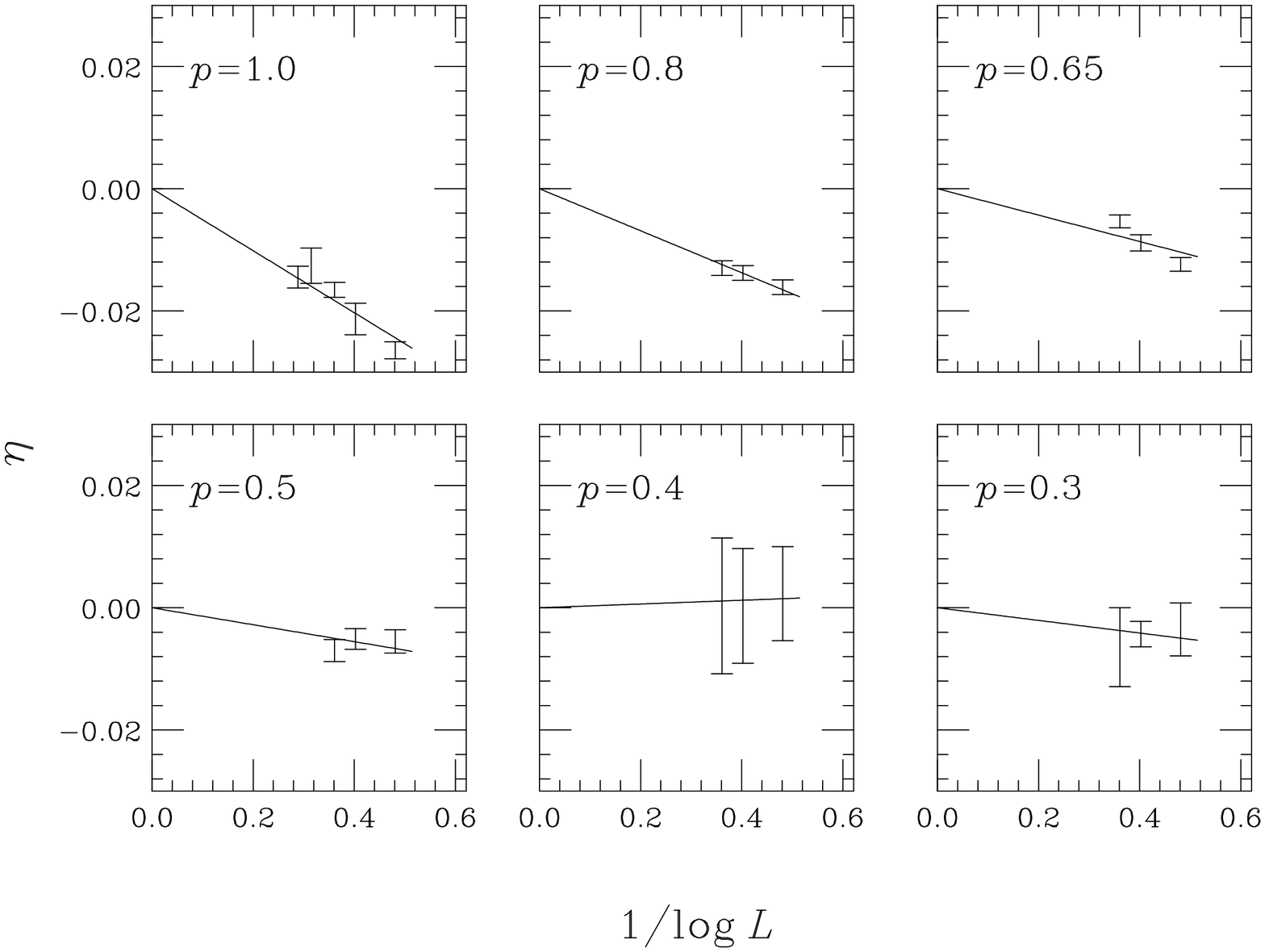,width=0.68\linewidth,angle=90}
\end{center}
\protect\caption{The $\eta$ exponent for different lattice sizes. 
The fit is enforced to yield $\eta=0$ for $L\to\infty$.
\protect\label{ETA1/logL}}
\end{figure}

As PRG predicts logarithmic corrections to the MF behavior, we should
expect scaling-corrections of order $1/\log L$ (for the $\eta$ and
$\nu$ exponents in the pure model and only for the $\eta$ exponent in
the diluted case) and $1/\sqrt{\log L}$ (for the $\nu$ exponent in the
diluted one): both are of the same order for the lattices that we can
afford!. In figures~\ref{NU1/logL} and \ref{ETA1/logL} we show that
the deviation from MF can indeed be accounted for by logarithmic
corrections. We have found similar results in two
dimensions~\cite{2D}. While finishing these papers, the same
conclusion has been independently drawn in a transfer-matrix study of
the bond-diluted two dimensional model~\cite{2Dbis}.

As we have seen, hyperscaling does not seem to hold. Indeed, the PRG
predict logarithmic violations. In this section, we try to identify
them by starting from a naive point of view. That is, we shall first
locate the critical point as if the usual FSSA were correct, and then we
shall study there the scaling of physical quantities, looking for
deviations from the pure-power law.

\begin{figure}[t]
\begin{center}
\leavevmode
{\centering\epsfig{file=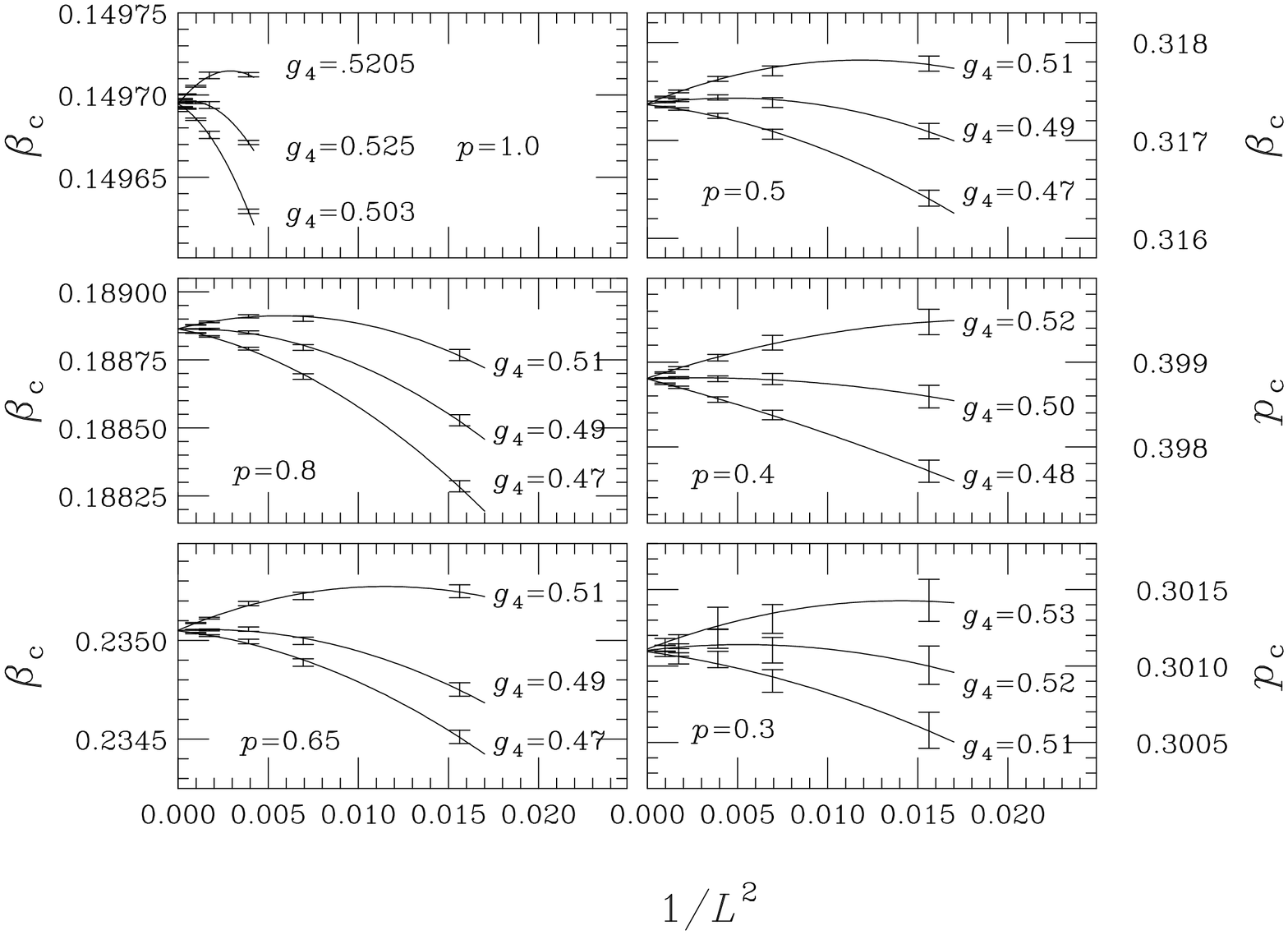,width=0.68\linewidth,angle=90}}
\end{center}
\protect\caption{{Computation of $\beta_\mathrm{c}$ or 
$p_\mathrm{c}$ looking at fixed 
values of $g_4$ for the different values of the dilution.
The lines are quadratic fits.}
\protect\label{CORTESVMFIJO}}
\end{figure}

A very accurate way of computing $\beta_\mathrm{c}$ (or
$p_\mathrm{c}$) is to fix a value of $g_4$, measuring in what $\beta$
the function $g_4(L,\beta)$ is equal to the fixed previous
value~\cite{PARUN}. 
In
absence of logarithmic corrections one expects that the shift in
$\beta$ behaves as $1/L^2$.  In figure \ref{CORTESVMFIJO} we plot
these quantities for three values of $g_4$ at each $p$, while the results
of a quadratic fit, for the infinite volume critical couplings are
found in table~\ref{BETAC}.  We point out that a logarithmic
correction as the one computed in ref.~\cite{KENNA} 
(as eq.~(\ref{FSSPURO}) shows) for the pure model 
or eq.~(\ref{shift}) for the diluted one, does not change the fit results.

In order to minimize the systematic errors, those values have been
obtained choosing a $g_4$ such that the linear coefficient in $1/L^2$
vanishes.  We have also computed the extrapolations for a wide range
of $g_4$ values to check the amplitude of the change.  We observe that, 
within one standard deviation in the extrapolated value, $g_4$ can be
changed in the interval $[0.3,0.8]$ in the best case ($p=0.8$) and in
the interval $[0.4,0.6]$ in the worst one ($p=0.5$).  Although we
think that systematic errors from this source are negligible, a more
conservative attitude could be to duplicate the statistical error bars in
table~\ref{BETAC}.

\begin{table}[t]
\begin{center}
\begin{tabular}{|l|l|}\hline
\multicolumn{1}{|c|}{$p_{\mathrm c}$}&
\multicolumn{1}{c|}{$\beta_{\mathrm c}$}\\\hline
1.0&0.149695(1)\\\hline
0.8 &0.188864(3)\\\hline
0.65&0.235049(8)\\\hline
0.50&0.317368(19)\\\hline
0.398806(18)&0.42\\\hline
0.30110(4)&0.633\\\hline
\end{tabular}
\caption{Critical couplings as obtained from a quadratic fit of
$\beta_c(L,g_4)$. The error bars are purely statistical.}
\label{BETAC}
\end{center}
\end{table}

Now, we can check hyperscaling violations of the form:

\begin{equation}
\xi(L,\beta_c)\propto L \left(\log L\right)^{\delta_\xi}\ .
\end{equation}

\begin{figure}[t]
\begin{center}
\leavevmode
{\centering\epsfig{file=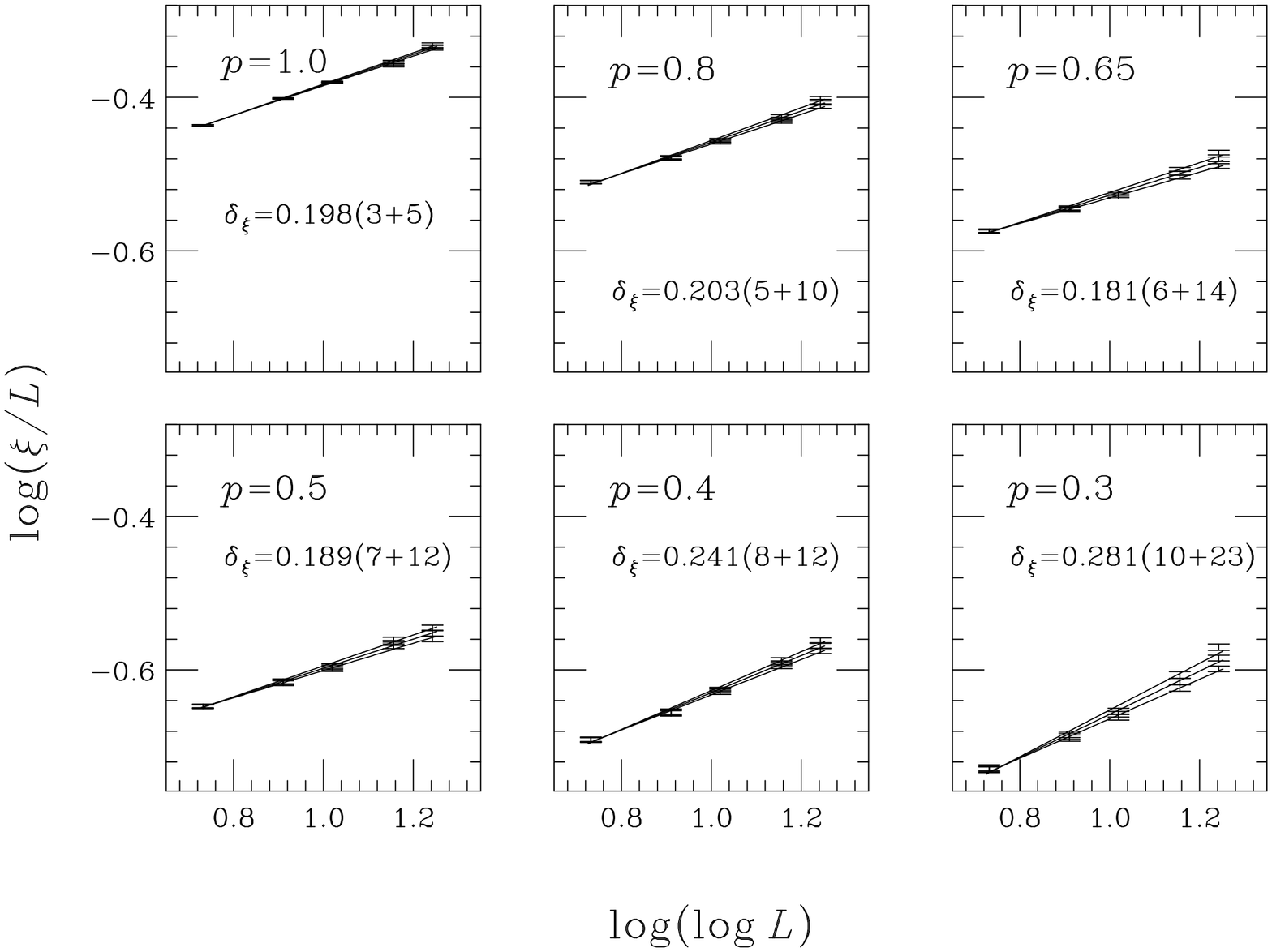,width=0.68\linewidth,angle=90}}
\end{center}
\protect\caption{{$\log (\xi/L)$ in the infinite volume critical point,
as a function of
$\log(\log L)$, for several dilutions. The fitted $\delta_\xi$ values are
also displayed. The first term in the error is statistical, while
the second corresponds to the error in the critical coupling.
The three lines correspond to data at the critical point and one
standard deviation apart at each side.}
\protect\label{HYPERSCALING}}
\end{figure}

In figure~\ref{HYPERSCALING}, we plot $\log (\xi/L)$ as a function of
$\log (\log L)$, the slope being directly $\delta_\xi$. We can see
that a good linear behavior is obtained. The fitted $\delta_\xi$-values
are reasonably close to the theoretical prediction, but a growing
trend with the dilution is self-evident. A naive (wrong) explanation
is that this is an effect coming from the vicinity of the percolation
critical point.  Indeed, we can repeat the previous fit in the pure
percolation for similarly-sized lattices. Fitting from $L=8$ we obtain
$\delta_\xi=0.0927(9)$ with $\chi^2/\mathrm{d.o.f.}=97.9$ and from
$L=16$, the fit parameters are $\delta_\xi=0.066(2)$ with
$\chi^2/\mathrm{d.o.f.}=6.90$. So a linear behavior is ruled out for
this model. However, this is not surprising, as hyperscaling is known
to hold for percolation in four dimensions.

Another interesting observable is the specific heat, which for the
pure model is expected to diverge as 
$C \propto \left(\log L\right)^{1/3}$,
while in the diluted case it is expected to remain bounded.
Both predictions can be tested. The fitted values for the logarithmic 
divergence exponent, $\delta_C$, are
\begin{equation}
\begin{array}{rcllrcl}
p&=&1.0&: \quad\qquad &\delta_C&=&0.399(4+22)\ ,\\
p&=&0.8&:        &\delta_C&=&0.304(7+13)\ ,\\
p&=&0.65\!\!\!&:        &\delta_C&=&0.184(8+15)\ ,\\
p&=&0.5&:        &\delta_C&=&0.095(6+9)\ ,\\
p&=&0.4&:        &\delta_C&=&0.084(5+6)\ ,\\
p&=&0.3&:        &\delta_C&=&0.073(8+7)\ .
\end{array}
\end{equation}

On the values of the Binder cumulants at the infinite volume critical
temperature we can  see directly the (logarithmic) corrections to the
scaling~\footnote{
        Do not confuse with the multiplicative logarithmic corrections to 
        the critical law that we have studied above.}. 
In the pure case we have obtained $g_4 \simeq 0.51$ for the largest
lattice at the critical point: we remark that the asymptotic value for
the pure model is $g_4 \simeq 0.406$. It is possible to show using~\cite{AMIT}
that the leading correction to scaling term for $g_4$ goes as $1/\log L$ in
both the pure model and the diluted one.
Our pure $g_4$-data are compatible
with a limiting value of $0.406$ modified by $1/\log L$ corrections.

In the diluted model, we have obtained data for the $g_4$
cumulant around $0.5$ (see fig.~\ref{CORTESVMFIJO}) which is quite
larger than the predicted value $g_4\simeq 0.32$ (see
eq.~(\ref{G4G2})). Again, this numerical discrepancy can be understood
taking into account corrections like $1/\log L$. The same comments hold
for $g_2$.  

\begin{figure}[t]
\begin{center}
\leavevmode
{\centering\epsfig{file=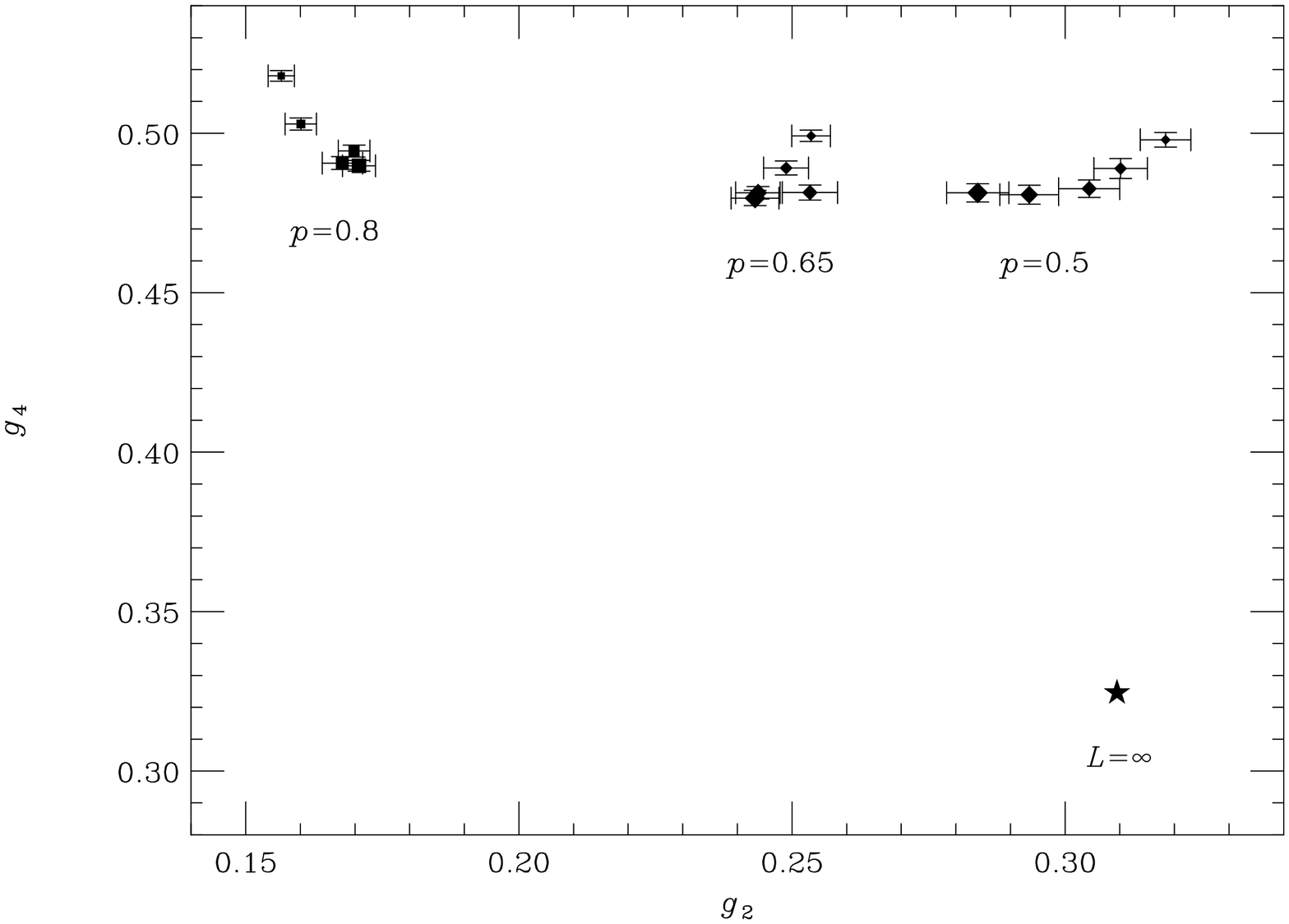,width=0.68\linewidth,angle=90}}
\end{center}
\protect\caption{{$g_4$ versus $g_2$ for three values 
of the dilution. The symbol sizes increase with the lattice size 
($L=8,12,16,24,32$). The PRG prediction is also plotted.
}
\protect\label{G2G4}}
\end{figure}

In figure ~\ref{G2G4} we have shown $g_2$ against $g_4$ for some
values of the dilution. We have also plotted the theoretical value for
the pair $(g_2,g_4)$. This figure must be interpreted having in mind
the previously cited logarithmic corrections to the scaling.  It is
clear that the complete numerical characterization of these cumulants
needs further numerical work. 

Finally we have studied the probability distribution of the observable
${\cal M}^2$ (i.e. we collect the histogram of the values of ${\cal
M}^2$ at every measure independently of the sample). We have found
that for large values of ${\cal M}^2$ the data follow a law $P({\cal
M^2}) \propto \exp(-c(L)({\cal M}^2)^2)$. Using the results of section
(\ref{S_BINDER}) it is possible to show that the theoretical
prediction for the probability distribution is a Gaussian with $c(L)
\propto u(L) L^4$. The coefficient, $c(L)$, computed numerically follows
very well a law $L^4/\sqrt{\log L}$ in perfect agreement with the
theoretical prediction.

\section{\protect\label{S_CON}Conclusions}

We have performed a Monte Carlo simulation of the site-diluted Ising
model in four dimensions, for several values of the dilution, and in a
wide range of lattice sizes. The use of a finite-size scaling analysis
allows us to consider big lattices just at the critical point. To gain
accuracy we have repeated the simulations for many different hole
configurations.

As a first stage, we have measured with great precision the critical
exponents under the hypothesis of hyperscaling. The value we obtain
for the $\nu$ exponent changes along the critical line, ruling out the
possibility of a single non-Gaussian fixed point.

Using Perturbative Renormalization Group techniques, we have computed
the scaling formulas for the diluted model, obtaining specific
logarithmic corrections to the Gaussian behavior.

We have re-analyzed our numerical data finding that they agree with a
Gaussian scenario with logarithmic corrections to hyperscaling. The
pure model has also been considered and its corresponding scaling
formulas, checked.

Although the nature of the logarithmic corrections hardly allows to
perform precise fits to the predicted functional forms, we have found
a reasonable agreement.

\section*{Acknowledgments}

We thank to the CICyT (contracts AEN94-0218, AEN96-1634) for partial
financial support, specially for the use of dedicated Pentium Pro
machines where we have carried out the simulations. JJRL is granted by
EC HMC (ERBFMBICT950429).

\hfill

\end{document}